\documentclass[preprint,showpacs,showkeys,aps]{revtex4}
\usepackage[dvips]{graphicx}% Include figure files
\begin{document}
\title{Nonequilibrium Temperature and Thermometry
in Heat-Conducting $\phi^4$ Models}
\author{
Wm. G. Hoover and Carol G. Hoover       \\
Ruby Valley Research Institute                  \\
Highway Contract 60, Box 598                    \\
Ruby Valley, Nevada 89833                       \\
}
\date{\today}
\pacs{05.20.-y, 05.45.-a,05.70.Ln, 07.05.Tp, 44.10.+i}
\keywords{Temperature, Thermometry, Thermostats, Fractals}

\begin{abstract}

We analyze temperature and thermometry for simple
nonequilibrium heat-conducting models.  We also show in
detail, for both two- and three-dimensional systems,
that the ideal gas thermometer corresponds to
the concept of a local instantaneous mechanical
kinetic temperature.  For the $\phi^4$ models
investigated here the mechanical temperature
closely approximates the local thermodynamic
equilibrium temperature.  There is a significant
difference between kinetic temperature and the
nonlocal configurational temperature.  Neither
obeys the predictions of extended irreversible
thermodynamics.  Overall, we find that kinetic
temperature, as modeled and imposed by the Nos\'e-Hoover
thermostats developed in 1984, provides the simplest means for
simulating, analyzing, and understanding
nonequilibrium heat flows.  

\end{abstract}

\maketitle

\section{Introduction}

The present work emphasizes and details the mechanical nature
of the kinetic temperature, in contrast to the ensemble-based
configurational temperature.  Simulations for the simple models
considered here are insensitive to system size and show
significant differences between the kinetic and configurational
temperatures.  Our main goal is to illustrate and emphasize the
relative advantages of kinetic temperature, particularly away
from equilibrium.

Ever since the early days of molecular dynamics ``temperature'' has
been based on the familiar ideal-gas kinetic-energy definition.
For a Cartesian degree of freedom at equilibrium the kinetic
definition is:
$$
kT_K \equiv \langle mv^2 \rangle \ .
$$
This definition provides a means for linking Gibbs' and Boltzmann's
classical statistical mechanics to thermodynamics.  Because
thermodynamic equilibrium corresponds to the Maxwell-Boltzmann
velocity distribution,
$$
f(v) = \sqrt{(m/2\pi kT)}\exp [-mv^2/2kT] \ ,
$$
{\em} any of the even moments, 
$$
\langle v^2 \rangle = 1 \times (kT/m) \ ; \
$$
$$
\langle v^4 \rangle = 1 \times 3 \times (kT/m)^2 \ ; \
$$
$$
\langle v^6 \rangle = 1 \times 3 \times 5 \times (kT/m)^3 \ ; \
$$
$$
\dots \ ,
$$
can be used to define the temperature for a system {\em at} equilibrium. 
The second-moment choice is not only the simplest, but in the ideal-gas
case it also corresponds to a conserved quantity, the energy.  The same
definition of temperature is a fully consistent choice {\em away} from
equilibrium too.

An {\em ideal-gas thermometer} can be visualized as a collection of
many very small, light, and
weakly-interacting particles, but with such a high collision rate
that thermal equilibrium (the Maxwell-Boltzmann distribution) is
{\em always} maintained within the thermometer.  For an innovative
implementation of this model with molecular dynamics see Ref. [1].

{\em Configurational} temperature definitions are also possible.  There
are two motivations for considering such coordinate-based temperatures:
first, there is some ambiguity in determining the mean velocity in a
transient inhomogeneous flow---kinetic temperature has to be measured
relative to the flow velocity while configurational temperature does
not; second,
the search for novelty.  The simplest of the many configurational
possiblities was suggested and also implemented by Jepps in his
thesis\cite{b2}.  In independent research, directed toward finding
a canonical-ensemble dynamics consistent with configurational temperature,
Travis and Braga developed an implementation identical to Jepps'
unpublished algorithm\cite{b3}.  The underlying expression for the
configurational temperature,
$$
kT_C \equiv \langle F^2 \rangle / \langle \nabla^2 {\cal H} \rangle \ ,
$$
appeared over 50 years ago in Landau and Lifshitz' statistical physics
text\cite{b4}.  In the definition of $kT_C$ the force $F$ for a
particular degree of freedom depends
upon the corresponding gradient of the Hamiltonian, 
$$
F = -\nabla {\cal H} \ .
$$
Landau and Lifshitz showed that the expression for $kT_C$ follows from
Gibbs' canonical distribution,
$$
f_{\rm Gibbs} \propto \exp [-{\cal H}/kT] \ ,
$$
by carrying out a single integration by parts:
$$
\langle \nabla ^2{\cal H} \rangle =
\langle (\nabla {\cal H})^2 \rangle / kT \longrightarrow
kT_C \equiv \langle (\nabla {\cal H})^2 \rangle / 
\langle \nabla ^2{\cal H} \rangle \ .
$$
Unlike the kinetic
temperature, the configurational temperature $T_C$ is not simply related
to a mechanical thermometer.  And in fact, there are {\em many} other
such nonmechanical temperature expressions.  Away from equilibrium it is
clear that no finite number of moments or averages can be expected to
uniquely define a phase-space distribution function.  For a thorough discussion
see Refs. [2] and [3].  Long before this complexity surfaced the proper
definition of temperature away from equilibrium was a lively subject.
To capture some of its flavor over a 30-year period see Refs. [5] and [6].
Relatively cumbersome microcanonical versions of configurational
temperature have been developed following Rugh's investigations.  For
references and an early application of these variants see
Morriss and Rondoni's work\cite{b7}.

Jou and his coworkers and their critics\cite{b8,b9,b10,b11,b12} have
considered the desirability of measuring an ``operational" ``thermodynamic"
temperature for nonequilibrium systems.  They discussed and then
implemented a method\cite{b8,b12} (which we explore in more detail here)
for its measurement.  Figure 1 illustrates the simplest case of their
idea, a heat conductor connected to a ``thermometer".  As usual, the
devil is in the details.  Here the details include both the {\em type} of
thermometer used and the linkage connecting that thermometer to the
conducting system.  The linkage certainly has an effect on the forces
and internal energy at the linkage point, and hence affects the
local-thermodynamic-equilibrium temperature and the configurational
temperature. In addition to their
``operational'' temperature, Jou {\it et alii} also consider a ``Langevin
temperature'', $T_{\rm Langevin}$ (the temperature which enters explicitly
into the usual equilibrium Langevin equations of motion) and a ``local
thermodynamic equilibrium'' temperature, $T_{\rm LTE}$ (the temperature
based on the equilibrium equation of state,
$$
T_{\rm LTE} \equiv T(\rho,e) \ ,
$$
where $e$ is the internal energy per unit mass).  {\em At} equilibrium,
and only at equilibrium, all of the various temperatures are the same and
there is no ambiguity in the temperature concept:
$$
T = T_K = T_C = T_{\rm Langevin} = T_{\rm LTE} \ {\rm [At \ Equilibrium]} \ .
$$
{\em Away from} equilibrium, where most physical interpretations of temperature
are actually symmetric second-rank tensors, we can expect that each of the
these four  ``temperatures'' differs from the others.  This {\em tensor} nature
of temperature is evident in strong shockwaves\cite{b13}.  Generally we
must anticipate that nonequilibrium temperature can be anisotropic, with
$$
T_{K,C,\rm{LTE}}^{xx} \ne T_{K,C,\rm{LTE}}^{yy} \ne T_{K,C,\rm{LTE}}^{zz} \ .
$$
This anisotropicity makes it imperative to describe the microscopic
mechanics of any nonequilibrium thermometer in detail and argues strongly
against a nonequilibrium version of the Zeroth Law of Thermodynamics.

In their illustrative
example, Jou and Hatano\cite{b12} used the temperature of a Langevin oscillator\cite{b14} coupled to a driven
oscillator to measure the driven oscillator's temperature.  A Langevin
oscillator is damped with a constant friction coefficient and driven with a
random force\cite{b14}.  See also the next-to-last paragraph of Sec. II.
Jou and Hatano\cite{b12} found that their measured
temperature was qualitatively sensitive to the assumed form of coupling
linking their ``system'' (the driven oscillator) to their ``thermometer''
(the Langevin oscillator).

At equilibrium, thermometry, and thermodynamics itself, both rely on
the observation often called the ``Zeroth Law of Thermodynamics'',
that two bodies in thermal equilibrium with a third are also in
thermal equilibrium with each other (independent of the couplings
linking the bodies).  Jou and Hatano drew the very reasonable
conclusion from their work that this fundamental property of temperature,
which makes equilibrium
thermometry possible, might be {\em impossible} away from equilibrium.

Baranyai\cite{b15,b16} considered a much more complicated thermometer,
a tiny crystallite, made up of a few hundred tightly-bound
miniparticles.  He compared both
the kinetic and the configurational temperatures of nonequilibrium flows
with the temperatures within his thermometer and found substantial
differences. Baranyai was able to conclude from his work that neither
the kinetic nor the configurational temperature was a ``good'' nonequilibrium
temperature.  By this, he meant that neither satisfied the Zeroth Law of
thermodynamics.
The temperature within Baranyai's minicrystal thermometer, his
``operational temperature'', exhibited relatively small spatial
variations (the entire many-body thermometer was about the same size
as a single particle of the nonequilibrium system in which it was immersed).

There is a considerable literature extending irreversible thermodynamics
away from equilibrium, based on defining the nonequilibrium temperature,
in terms of an (ill-defined) nonequilibrium entropy:
$$
T = (\partial E/\partial S_{\rm noneq})_V \ .
$$
For a recent guide to the literature see Ref. [17].

{\em At} equilibrium, Gibbs and Boltzmann showed that the entropy $S_{\rm eq}$
of a classical system is simply the averaged logarithm of the phase-space
probability density,
$$
S_{\rm eq} = -k\langle \ln f_{\rm eq} \rangle \ .
$$
{\em Away from} equilibrium $f_{\rm noneq}$ is typically fractal\cite{b18,b19}
(so that its logarithm diverges), so that the very existence of a
{\em nonequilibrium} entropy appears doubtful. For a comprehensive review
of efforts based on a {\em nonequilibrium} Gibbs entropy, presumably
$-k\langle \ln f_{\rm noneq} \rangle $,
see Ref. [20]. It is
evident that such efforts are inconsistent with what is known about the
singular fractal nature of nonequilibrium phase space distributions,
$\{ \ f_{\rm noneq} \ \}$.

Recent thorough work by Daivis\cite{b21} investigated the consequences of an
{\em assumed} nonequilibrium entropy.  Daivis compared three equalities
(analogous to the equilibrium Maxwell Relations) based on the assumed
existence of $S_{\rm noneq}$ with results from numerical simulations.
{\em None} of the three ``equalities" was satisfied by the simulation
results, casting doubt on both the existence of a nonequilibrium entropy
analogous to the Gibbs-Boltzmann entropy, and also on the existence of a
corresponding entropy-based temperature.

In the present work we will explore these ideas for a simple
nonequilibrium model of heat flow, the $\phi ^4$ model\cite{b19,b22}.
This very basic model has quadratic Hooke's-law interactions linking nearest
neighbor pairs of particles.  In addition, each particle is tethered to
its individual lattice site with a quartic potential.  This model has
been extremely useful for nonequilibrium statistical mechanics.  In its
most useful temperature range [where the particles are sufficiently
localized, as detailed in Sec. IV] we will see that the internal energy varies
nearly linearly with kinetic temperature, simplifying analyses.  The
model obeys Fourier's law (for small enough temperature gradients for
the equivalence of all the various temperature definitions), even in
one dimension\cite{b22}.  It can also display considerable phase-space
dimensionality loss\cite{b19}, establishing the
fractal nature of the phase-space distribution function.  Because the
loss can exceed the phase-space dimensionality associated with the
thermostating particles, a {\em fractal} distribution for the
interior Newtonian part of a driven nonequilibrium system is implied
by these results.  We use the $\phi ^4$ model here to elucidate and
compare the kinetic and configurational candidates for nonequilibrium
temperature.

Though the mechanical models we consider are small, with only a few dozen
degrees of freedom, we firmly believe that the analysis of such very specific
manageable models is the only reliable guide to an understanding of
thermometry and temperature.  The pitfalls and complexities associated with
large systems, 
and with large thermometers, are the gradients and inhomogeneities already
seen in Baranyai's work\cite{b15,b16}.

The paper is organized as follows: first, a discussion of mechanical
thermometry, using the ideal-gas thermometer, with simulations
corresponding to ideal gases of disks (two dimensions) and spheres
(three dimensions); next, a description of
the computer experiment suggested by Jou as applied to the $\phi ^4$
model.  After discussing and illustrating the $\phi ^4$ model, numerical
results, and conclusions based on them, make up the final sections of
this work.

%FIGURE 1
\begin{figure}
\includegraphics[height=6cm,width=6cm,angle=-90]{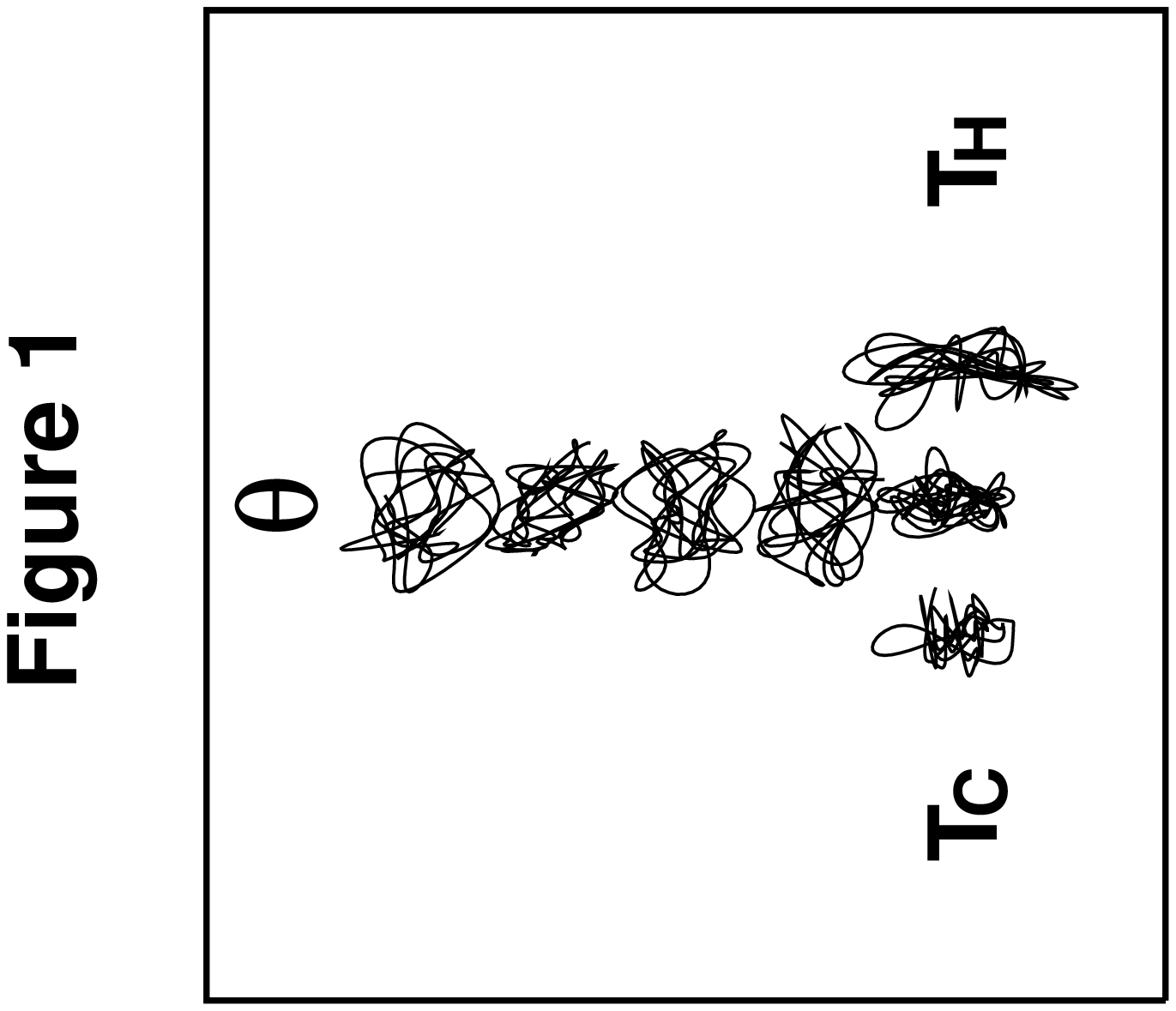}
\caption{
Jou's nonequilibrium system (described in detail in Sec. III),
driven by the temperature difference
$T_H - T_C$, is coupled to a thermometer which reads the ``actual''
or ``correct'' or ``equilibrium'' or ``operational'' temperature
$ T_\theta $.  This idea underlies our own simulations.   Here $ T_\theta$
represents temperature ``at'' the contact point between the
vertical ``thermometer'' and the particle located between the two
thermostated particles.  Each of the seven particles in the system is
represented here by a short trajectory fragment.
}
\end{figure}

\section{Ideal Gas Thermometry}

Hoover, Holian, and Posch\cite{b9} described the mechanics of a
one-dimensional ideal-gas thermometer in detail.  They considered
a massive particle, with momentum $MV$, interacting with a
Maxwell-Boltzmann bath of  ideal-gas particles with momenta
$\{ mv \}$.  Here we will consider the same situation in detail for
two- and three-dimensional thermometers.  A typical collision
can be viewed in the center-of-mass frame, a coordinate frame with
the center-of-mass velocity:
$$
v_{\rm com} = \frac{MV + mv}{M + m} \ .
$$
For an instantaneous hard-sphere impulsive collision the direction
of the {\em relative}
velocities in this frame, averaged over all possible collisions of the
two velocities,
$$
\{ v_{\rm before} \} = \pm(V - v) \ ,
$$
is directed {\em randomly} after collision.  This simplification leads to
a systematic expansion\cite{b9} of the energy change of the massive particle
in half-integral powers of the mass ratio $m/M$.  To second order in
$\sqrt{\frac{m}{M}}$:
$$
\langle (d/dt)(MV^2/2) \rangle \propto
(MV^2/2) - \langle (mv^2/2) \rangle = (MV^2/2) - (3kT_K/2) \ ,
$$
where $T_K$ is the ideal gas kinetic temperature.

For the details of other models (soft spheres, square wells, ...) of
the interaction between the massive
particle and an
ideal-gas-thermometer heat bath a solution of the corresponding
Boltzmann equation would be required.  Nevertheless, on physical
grounds it is ``obvious" that a massive particle with
(above/below)-average energy will (lose/gain) energy, on the average,
as a result of its collisions with the equilibrating bath,
$$
\langle \dot E \rangle \sim {\rm sign} ( \langle E \rangle_{\rm eq} - E) \ .
$$

It is an interesting exercise in numerical kinetic theory to confirm
this expectation in two and three dimensions.  Consider first a hard
disk with unit radius and mass $M$ with unit velocity $V = (1,0)$.
Scattering for disks is anisotropic.  On the average a disk retains
a memory of its original velocity in the center-of-mass frame.
To model the interaction of a massive disk with a heat bath of
unit-mass-point particles at kinetic temperature $T_K$ requires choosing
Maxwell-Boltzmann bath-particle velocities $\{ v \} = \{v_x,v_y\}$
as well as an angle
$0<\alpha < 2\pi$ for each collision, which specifies the location of
the colliding bath particle
relative to the massive disk.  See Fig. 2 for typical results.  These
were obtained by using a random number generator\cite{b23} to simulate
the collisions.

The velocity changes of the disk, $\Delta V$, and the bath particle,
$\Delta v$, are as follows for a collision described by the angle
$\alpha $:
$$
\Delta V = (V-v)\cdot(R-r)(\cos (\alpha),\sin (\alpha))[2m/(M+m)] \ ; \
$$
$$
\Delta v = (v-V)\cdot(R-r)(\cos (\alpha),\sin (\alpha))[2M/(M+m)] \ .
$$
A sufficiently long series of velocity changes $\{ \Delta V\}$,
computed in this way, can be used to find the averaged hard-disk
energy change shown in the figure.

Results for $m=1;M=100$ and five million randomly-chosen hard-disk
collisions for each ideal-gas temperature are shown in Fig. 2.  In
analyzing these simulations it is necessary
to weight the summed-up contributions of all the observed collisions
with the relative velocities corresponding to each collison $c$:
$$
\langle \Delta E \rangle =
\frac{\sum (|v-V|\Delta E)_c}{\sum (|v-V|)_c} \ .
$$
The speed $|v-V|$ is included because the collision rate for two
randomly located particles with velocities $v$ and $V$ is directly
proportional to the magnitude of their relative velocity, $v-V$.

As expected, the temperature at which the disk kinetic energy, for $M$
equal to one hundred, is equal to the averaged mass-point
thermal energy, is 50:
$$
\langle \Delta\frac{MV^2}{2} \rangle \propto 2kT_{\rm bath} - MV^2 \ .
$$
The analogous averaged mass-point thermal energy is $33\frac{1}{3}$ 
for hard spheres.  Energy changes for both disks and spheres are
shown in Fig. 2. The simplicity of such a mechanical model for a
thermometer---which ``measures temperature'' in terms of the kinetic
energy per particle---recommends its use in analyzing nonequilibrium
simulations.

The {\em configurational} temperature, on the other hand, has no
corresponding mechanical model, and also requires that the
quotient of {\em two} separate averages be computed to find the
temperature associated with a particular Cartesian degree of freedom,
$$
kT_C \equiv \frac{\langle F^2 \rangle }
{\langle \nabla^2 {\cal H} \rangle} \ .
$$
{\em Kinetic} temperature is simpler, requiring only a single average
because $\nabla _p^2 {\cal H} = 1/m$ is constant:
$$
kT_K \equiv \langle (\nabla _p{\cal H})^2 \rangle / 
\langle (\nabla ^2_p{\cal H}) \rangle = \langle p^2 \rangle /m = 
m \langle v^2 \rangle \ .
$$
Unlike the kinetic temperature the configurational temperature is
nonlocal (through its dependence on forces).

It should be noted that the ``Langevin thermometer'', as implemented
by Hatano and Jou\cite{b12}, {\em appears} to be based on a similar \
application of kinetic theory.  But the Langevin thermometer, if viewed
as a ``thermostat'' designed to impose the temperature $T_{\rm Langevin}$
suffers from the defect that its ``temperature'' (given by the ratio of
the time-integrated correlation function of the fluctuating force
to the drag coefficient) is {\em not} equal to
$\langle mv^2/k\rangle $ (or to any other oscillator-based temperature)
except {\em at} equilibrium.  
The ideal-gas thermometer, on the other hand, maintains its
temperature both at and away from equilibrium, and can easily be
implemented in numerical simulations by using either Gaussian
(constant kinetic energy) or Nos\'e-Hoover (specified time-averaged
kinetic energy) mechanics.  Both these thermostats employ feedback
forces to maintain the specified kinetic temperature $T_K$ even away
from equilibrium.

Baranyai's thermometer\cite{b15,b16}, with hundred of degrees of freedom,
contains within it both stress and temperature gradients.  His
minicrystal thermometer translates, rotates and vibrates as well.
This complexity destroys the local instantaneous nature of temperature
that is so valuable for analyzing inhomogenous systems with large
gradients.

%FIGURE 2
\begin{figure}
\includegraphics[height=6cm,width=6cm,angle=-90]{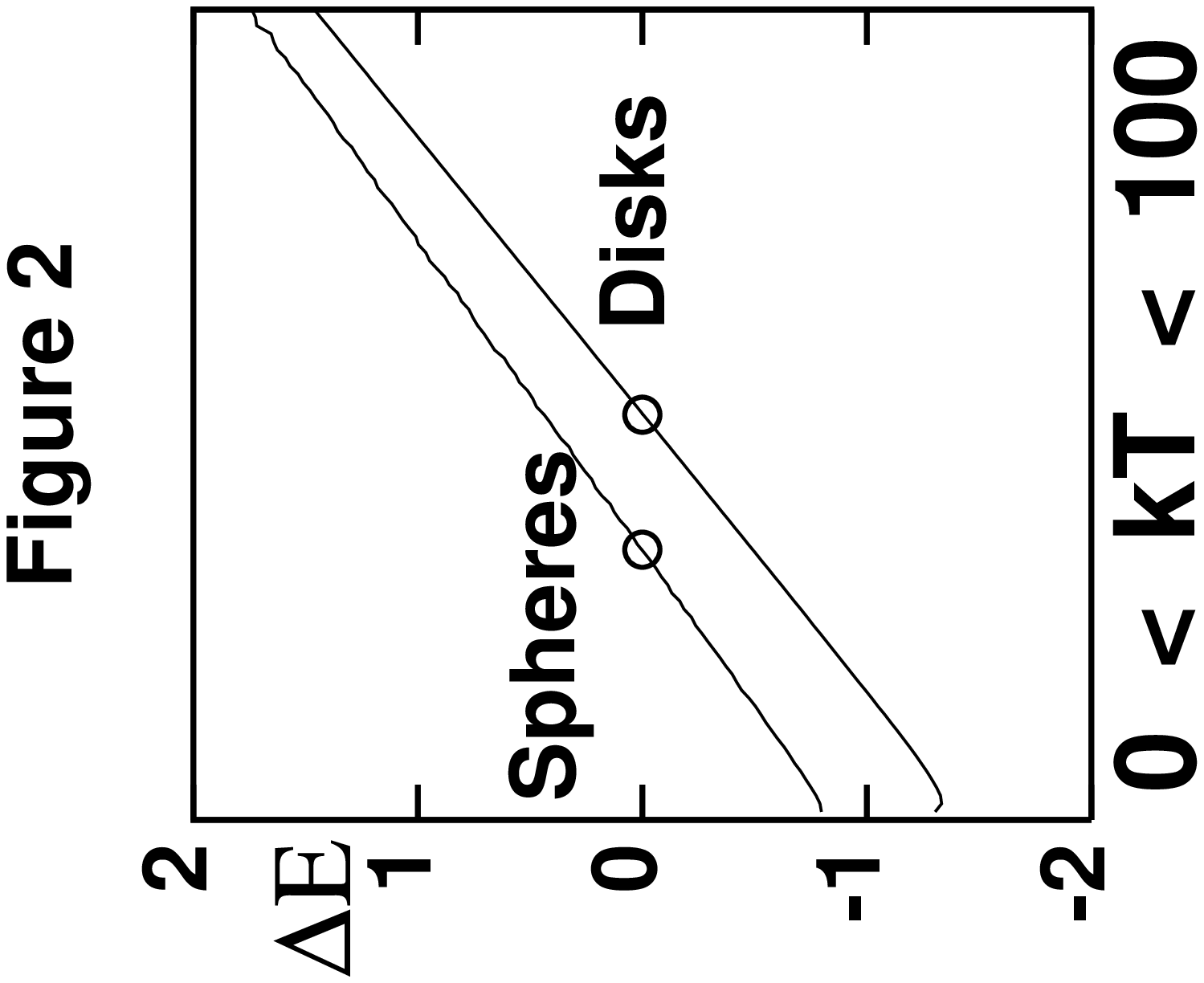}
\caption{
Energy change, due to collisions, for a hard disk of mass $M$ and unit
speed with an equilibrium bath of point particles with mass
$m = M/100$  and temperature $T_K$.  Zero energy change corresponds
precisely to that temperature (50 for disks, $33\frac{1}{3}$ for spheres;
open circles in the figure)
 for which the disk kinetic energy equals
the mean bath energy, $\langle mv^2/k \rangle $.  Analogous results
for a hard sphere immersed in a hard-sphere ideal-gas thermometer
are shown too.
}
\end{figure}

\section{Jou's Thermometric Experiment}

In order to explore the concept of nonequilibrium temperature Jou
suggested\cite{b8}, and ultimately tested\cite{b12}, the setup
shown in Fig. 1.  As indicated in that figure, an
equilibrium thermometer measures the ``real'', or ``thermodynamic'',
or ``operational'' temperature $ T_\theta $ when it is connected to a
nonequilibrium system with a temperature intermediate to $T_{\rm hot}$ and
$T_{\rm cold}$.  The constraint on individual particles' velocities
imposed by the heat current in the nonequilibrium system suggests that
the nonequilibrium temperature  $ T_\theta $ will turn out to be lower
than the local thermodynamic
equilibrium temperature $T_{\rm LTE}$ (the temperature based on mass,
momentum, and energy through the equilibrium equation of state)\cite{b8,b11}.
``Extended Irreversible Thermodynamics''\cite{b17} provides an estimate
for this temperature difference:
$$
T_{\rm LTE} - T_\theta \simeq Q^2 \ ,
$$
where $Q$ is the heat flux and the proportionality constant in this
relation is a temperature-and-density-dependent material property.
Although Hatano and Jou\cite{b12} confirmed that the kinetic temperature for a
simple two-oscillator model actually {\em is} less than the temperature
measured by a Langevin thermometer, the configurational temperature for
this same model behaved oppositely, {\em exceeding} the Langevin
temperature!  This discrepancy led Hatano and Jou to conclude that
the Zeroth Law of thermodynamics is unlikely to be obeyed away from
equilibrium, once again shedding doubt on the existence of a
nonequilibrium entropy.

In the present work we implement an extension of the Hatano and Jou
simulation to a two-dimensional few-body system based on the
$\phi ^4$ model\cite{b19,b22}, as described in the following section.

\section{$\phi ^4$ Model for Nonequilibrium Thermometry}

We consider a simple heat conducting nonequilibrium system in two
space dimensions.
See Fig. 3 for a time exposure of the corresponding dynamics.
There is a cold particle obeying the Nos\'e-Hoover equations of
motion:
$$
\dot x = (p_x/m) \ ; \
\dot y = (p_y/m) \ ; \
$$
$$
\dot p_x = F_x - \zeta _{\rm cold}p_x \ ; \
\dot p_y = F_y - \zeta _{\rm cold}p_y \ ; \
$$
$$
\dot \zeta _{\rm cold} \propto (p_x^2 + p_y^2 - 2mkT_{\rm cold}) \ .
$$
Both the cold particle and an analogous hot particle (with $\zeta _{\rm hot}$
and $T _{\rm hot}$) are connected to a Newtonian particle with
quadratic nearest-neighbor Hooke's-law bonds:
$$
\phi _{\rm Hooke} = \frac{\kappa _2}{2}(r-d)^2 \ .
$$
See again Fig. 3.
%FIGURE 3
\begin{figure}
\includegraphics[height=6cm,width=6cm,angle=-90]{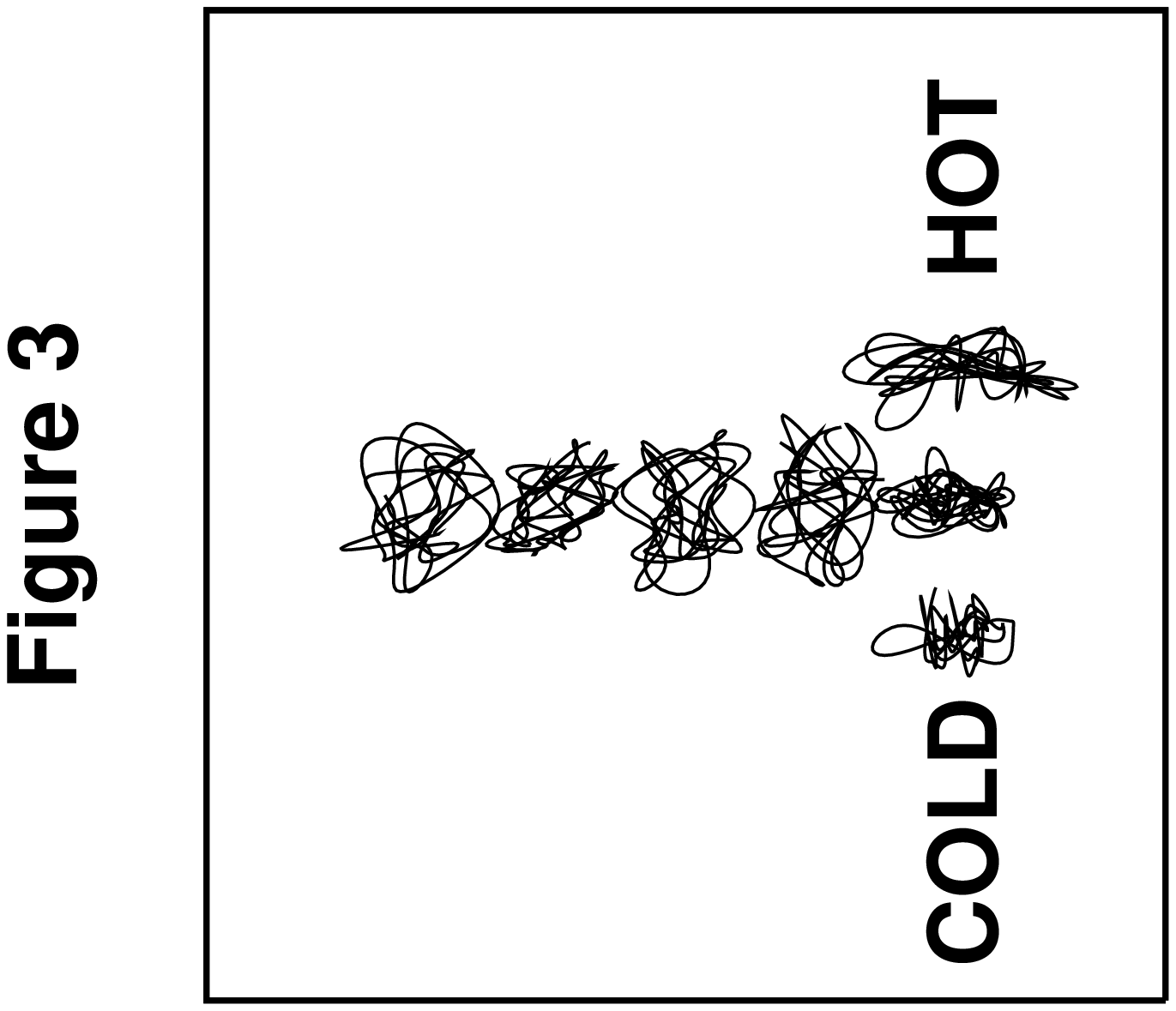}
\caption{
Particle trajectories for 20,000 timesteps.  The cold particle
kinetic temperature, $T_K^{\rm cold} = 0.01$, and the hot particle
kinetic temperature, $T_K^{\rm hot} =0.03$, are constrained with
Nos\'e-Hoover friction coefficients.  The corresponding measured
configurational temperatures are 0.0159 and 0.0265.  The longtime
averaged kinetic and configurational temperatures
of the five Newtonian particles are (from bottom to top):
$\{0.0207,0.0237,0.0238,0.0238,0.0242\}$ and
$\{0.0218,0.0229,0.0229,0.0229,0.0234\}$ respectively.  See Table
II.  The heat flux is
0.00269.}
\end{figure}

The Newtonian particle through which the flux $Q$ flows, from the hot
particle to the cold one on the average, lies at the end of a chain of similar
Newtonian particles.  This chain of Newtonian particles acts as
a {\em thermometer} through which no heat flows.  

To validate the chain idea we carried out preliminary {\em equilibrium}
simulations, with the ``hot'' and ``cold'' particles thermostated at a
common temperature:  $T_K^c = T_K^h = 0.07$.  Simulations with $10^9$
timesteps (beginning after first discarding half a billion equilibration
timesteps) were carried out for seven-, fourteen-, and 21-particle
systems.  These three simulations each provided time-averaged
configurational and kinetic temperatures for {\em all} particles lying in
the range $(0.0698 < T < 0.0701)$.  These simulations indicated consistent
equilibration
along the chains and between the configurational and kinetic temperatures
within a reasonable tolerance, $\pm 0.0001$.
We conclude from these equilibration runs that the $\phi ^4$ model is a
sufficiently mixing and conducting system for use in {\em nonequilibrium}
thermometry simulations.

This convincing equilibration suggests that a {\em chain} of $\phi^4$
particles {\em is} a suitable thermometer.  How long should the chain
be away from equilibrium?  To find this out we next carried out an
exactly similar series of three {\em nonequilibrium} simulations with
an extreme factor-of-19 difference between the constrained cold and
hot kinetic temperatures,
$$
T_K^c = 0.005 \ ; T_K^h = 0.095 \ .
$$
The long-time-averaged temperature results for seven-,
fourteen-, and 21-particle systems, shown in Fig. 4,  are essentially
the same, so that
a simple four-particle chain of thermometric particles is sufficient.

%FIGURE 4
\begin{figure}
\includegraphics[height=10cm,width=6cm,angle=-90]{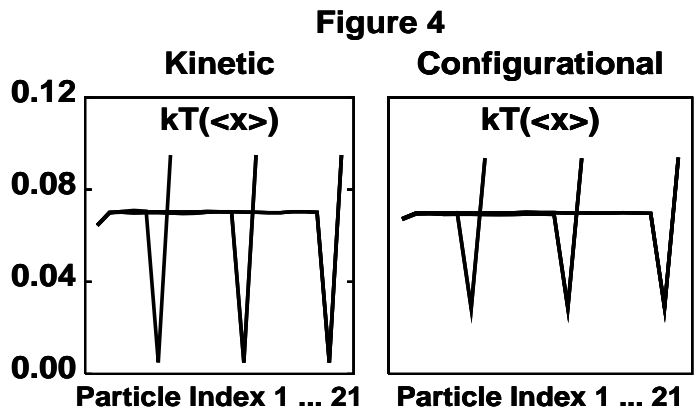}
\caption{
Long-time-averaged temperature profiles for nonequilibrium systems
of $n = \{ 7,14,21 \} $ particles.  Nos\'e-Hoover kinetic constraints
control the kinetic temperatures of a ``cold'' particle, with $T_K^c=0.005$,
Particle $n-1$, and a ``hot'' particle, with $T_K^h=0.095$, Particle $n$.
Particle 1 lies between the ``cold''
particle and the ``hot'' particle.  Both the kinetic and the configurational
temperatures are shown for all $n$ particles.  These simulations used
one billion timesteps after
discarding an equilibration run of half a billion timesteps.  $dt = 0.005$.
}
\end{figure}

Each of the particles
in this nonequilibrium system is tethered
to its lattice site $r_0$ with a quartic potential:
$$
\phi _{\rm Tether} = \frac{\kappa _4}{4}(r-r_0)^4 \ .
$$
With seven particles there are sixteen ordinary differential equations to
solve (seven coordinates, seven momenta, and two friction
coefficients).
For convenience we choose all of the particle masses, Boltzmann's
constant $k$, the force constants $\kappa _2$ and $\kappa _4$, the
Hooke's-Law equilibrium spacing $d$, and the cold and hot
proportionality constants determining the Nos\'e-Hoover
friction coefficients, all equal to unity.  For
the cold particle we solve the following equations:
$$
\dot x = p_x \ ; \
\dot y = p_y \ ; \
$$
$$
\dot p_x = F_x - \zeta _{\rm cold}p_x \ ; \
\dot p_y = F_y - \zeta _{\rm cold}p_y \ ; \
$$
$$
\dot \zeta _{\rm cold} = (p_x^2 + p_y^2 - 2T_{\rm cold}) \ . \
$$
We have carried out many other simulations, using configurational or
one configurational and one kinetic thermostat, as well as
different particle numbers, but the results are qualitatively
similar to those obtained with kinetic thermostats and are
therefore not reported here.  Likewise we do not explicitly consider here
the possibility of separately thermostating the $x$ and $y$ directions
(by using two friction coefficients, $\zeta _T^{xx}$ and $\zeta _T^{yy}$).

It should be noted that the Hooke's-Law nearest-neighbor potential
leads to {\em discontinuous forces} whenever particle trajectories
{\em cross} one another.  This is a common occurence in either one or two
dimensions, at sufficiently high temperatures.  In one or two dimensions
the force changes from $\pm 1$ to $\mp 1$ as two particles pass
through one another.  To avoid (or at least minimize) these
discontinuities in the present two-dimensional simulations we have
only considered simuilations with average temperatures less than or
equal to 0.1.

In discussing the applicability of irreversible thermodynamics to
nonequilibrium systems several workers have suggested the use of
a ``local thermodynamic equilibrium''
temperature\cite{b5,b8,b11,b17,b20,b24}.  For
the present model the relation between the local thermodynamic equilibrium
temperature and the kinetic temperature is nearly linear.
Figure 5 shows the variation of kinetic energy with internal energy
for a periodic chain of 100 particles (results for seven- and
fourteen-particle chains are essentially the same).  To an accuracy
better than a percent
$$
T_K \propto T_{\rm LTE} \ .
$$

%FIGURE 5
\begin{figure}
\includegraphics[height=6cm,width=6cm,angle=-90]{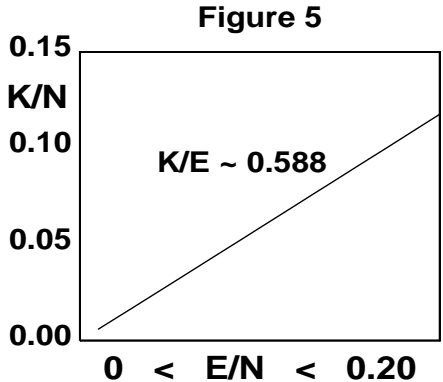}
\caption{
Variation of kinetic energy with total energy for a 100-particle
$\phi^4$ chain at equilibrium.  For each of the 20 points which
the line connects here $10^7$ timesteps
were used after discarding $5 \times 10^6$ equilibration timesteps.
$dt = 0.005$. To an excellent approximation $K \simeq 0.588E$.
}
\end{figure}

\section{Numerical Results and Concluding Remarks}

Exploratory simulations of the type illustrated in Figs. 3 and 4
suggested that
the kinetic and configurational temperatures are a bit different
(away from equilibrium) and also that these temperatures vary
slightly along the length of the Newtonian thermometric chain.  At
the same time the heat flow between the hot and cold particles closely
follows Fourier's law.  To show this explicitly Table I gives the
kinetic and configurational temperatures for an average temperature,
$T^{\rm av} = (T^c + T^h)/2 = 0.05$ and a broad range of temperature
differences, $\Delta T = T^h - T^c$.

The tabulated results for temperature differences which are not too
large,
$$
\Delta T/T^{\rm av} < 1 \ ,
$$
show a relatively small variation of the effective thermal
conductivity for the three-particle (cold-Newton-hot) system,
$$
\kappa = 2Q/(T_K^h - T_K^c) \ ,
$$
with the imposed temperature gradient.  There are significant differences
between the (local) kinetic and (nonlocal) configurational temperatures
of the two thermostated particles.  Similarly, the kinetic and
configurational temperatures
of the Newtonian particle linking them also differ somewhat.  On
the other hand, the near proportionality of
the internal energy and the kinetic energy {\em at equilibrium}
implies that local-thermodynamic-equilibrium temperature profiles
and kinetic-temperature profiles are essentially the same.

In every case the difference between the temperature of the Newtonian
particle {\em with} a heat flux (Particle 1) and the temperatures of the
thermometric Newtonian particles without a heat flux (Particles 2...5)
is rather small, but significant.  This difference is explored
systematically in Table II, where a relatively large kinetic temperature
difference,
$$
T_K^h = 3T_K^c \ \rightarrow \Delta T/T^{\rm av}  = 1 \ ,
$$
is imposed.  Symmetry suggests that the temperature difference should
depend quadratically on the heat flux (this same dependence is also
predicted by ``extended irreversible
thermodynamics''\cite{b8,b9,b10,b11,b17,b20,b24}). These simple
arguments are wrong.  In fact, the data in Table II
suggest a square-root
rather than a quadratic dependence.  Figure 6 shows the dependence of
the temperature differences $T^5_K-T^1_K$ and $T^5_C-T^1_C$ on the
heat flux $Q$.

The data in both Tables, calculated with all the Hooke's-Law force
constants equal to unity, are consistent with the set of
{\em nonequilibrium} inequalities:
$$
T_\theta > T_C > T_K \ ,
$$
where $T_\theta$ is the thermometric temperature of the Newtonian
thermometer while  $T_C$ and $T_K$ are the configurational and kinetic
temperatures of the Newtonian particle through which heat flows.  On
the other hand, simply reducing the force constant (from 1.0 to 0.3)
linking that Newtonian particle to the thermometric chain (and leaving
all the other force constants unchanged) gives {\em different} inequalities:
$$
T_C > T_\theta > T_K \ .
$$
Whether or not the conducting Newtonian particle is ``hotter'' or ``colder''
than the thermometric chain depends on the definition of temperature
{\em at} that particle!  The anistropicity of the Newtonian particle's
temperature is relatively small in these simulations, and tends to decrease
as the force constant linking that particle to the thermometric chain
is decreased.  For instance, $T_K^{yy} - T_K^{xx}$ is reduced from
0.012 to 0.006 as the linking force constant is reduced from 1.0 to 0.1.
The sign of this disparity,
$T_K^{yy} > T_K^{xx}$, is nicely consistent with the intuitive reasoning
of Jou and Casas-V\'asquez\cite{b8,b11}.

Evidently the predictions of extended irreversible thermodynamics
are not particularly useful in understanding the temperature differences
which result from small-system thermometry with relatively large
thermal gradients.  The detailed results depend
upon the details of the thermometric linkage.  Note that
the configurational temperature of the hot/cold thermostated particle
lies {\em below/above} the kinetic temperature, a symptom of the
configurational temperatures' nonlocality.  Because the {\em sign} of
$T_K - T_C$ can vary, both mechanical and thermodynamical effects are
involved.

In order to show that the qualitative features of thermometry for
the $\phi^4$ model are insensitive to temperature, we collect typical
results in Table II for sets of cold and hot
temperatures varying over two orders of magnitude.  In each case the
kinetic temperatures of the cold and hot particles are imposed by
Nos\'e-Hoover thermostats.  Then the long-time-averaged temperatures,
both kinetic and configurational, are measured for all of the particles.
The averaged heat flux is included too.  The
configurational temperature of the ``cold'' particle is uniformly
higher than its kinetic temperature, while the configurational temperature
of the ``hot'' particle is uniformly lower.  This
complexity is due to the nonlocal character of configurational
temperature.

In summary, let us reiterate our findings.  First, numerical kinetic
theory simulations (Fig. 2) demonstrate the local instantaneous dynamical basis
of kinetic temperature.  Next, stationary heat flows demonstrate an
insensitivity of the nonequilibrium temperature to system size (Fig. 4)
and also show that the kinetic and configurational temperature shifts
away from equilibrium can differ by more than a factor of two.  This
disparity occurs despite the near equivalence (Fig. 5) of the kinetic
temperature to the local-thermodynamic equilibrium temperature.  Though
it is possible to imagine and compute many ``temperatures'' away from
equilibrium, none of which satisfies a Zeroth Law, we see
no reason to prefer any definition more complicated than that
of the ideal-gas thermometer.  A mechanical, local, and instantaneous
physical thermometer
(which also corresponds well to a local thermodynamic equilibrium
thermometer in the present case) is appealing.  It is the simplest choice.

A particularly interesting problem where locality is important
for nonequilibrium thermometry is the stationary shockwave.  There
the differences between the longitudinal and transverse kinetic
temperatures are extremely large (as measured by an ideal-gas
thermometer) and the relaxation times are determined by the atomic
vibration frequency rather than diffusive processes\cite{b13}.  The
extreme spatial gradients associated with strong shockwaves make
the smoothing associated with configurational temperature undesirable.

%FIGURE 6
\begin{figure}
\includegraphics[height=6cm,width=6cm,angle=-90]{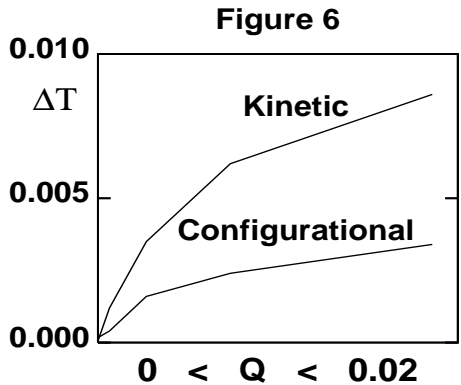}
\caption{
Variation of the kinetic-temperature and configurational-temperature
differences
with heat flux, using the data from Table II.  A {\em quadratic} variation
on this plot (rather than the apparent {\em square root}) corresponds to
the ``predictions'' of extended irreversible thermodynamics.
}
\end{figure}

\begin{acknowledgments}

Peter Daivis, Leopoldo Garc\'ia-Col\'in, Janka Petravic, Ian Snook,
Billy Todd, Karl Travis, and Paco Uribe all made comments and
suggestions as this work progressed.  Karl's, Peter's, and Janka's generous
suggestions and corrections were specially useful.  Billy's kind
invitation to visit Melbourne for MM2007 and Ian's hospitality at
the Royal Melbourne Institute of Technology made this work possible.
We thank one of the referees for pointing out that the concept of
nonequilibrium temperature, emphasizing anisotropicity, is exhaustively
reviewed in J. Casas-V\'asquez and D. Jou, Rep. Prog. Phys. {\bf 66},
1937 (2003).

\end{acknowledgments}

\newpage
\newpage

Table I.  Averages for runs of length $t = 5,000,000$ with the
fourth-order Runge-Kutta timestep $dt = 0.005$.
The kinetic and configurational temperatures are listed, along with
the heat flux $Q$ (all accurate to the last figure).  The first
seven columns correspond to the temperatures of the cold and hot
particles, followed by the temperature of the Newtonian particles
(the Newtonian particles are the five shown in a vertical column
in Fig. 3, and labeled from bottom to top.)
\begin{table}[th] 
\begin{tabular}{cccccccc}
\hline
$T_K^c$ & $T_K^h$ & $T_K^1$ & $T_K^2$ & $T_K^3$ & $T_K^4$ & $T_K^5$ & $Q$ \\   
\hline
0.045 & 0.055 & 0.0504 & 0.0507 & 0.0507 & 0.0507 & 0.0507 & 0.0020\\ 
0.040 & 0.060 & 0.0512 & 0.0524 & 0.0526 & 0.0526 & 0.0528 & 0.0039\\ 
0.035 & 0.065 & 0.0526 & 0.0554 & 0.0558 & 0.0559 & 0.0560 & 0.0057\\ 
0.030 & 0.070 & 0.0542 & 0.0588 & 0.0593 & 0.0594 & 0.0595 & 0.0076\\ 
0.025 & 0.075 & 0.0559 & 0.0622 & 0.0628 & 0.0629 & 0.0631 & 0.0094\\ 
0.020 & 0.080 & 0.0574 & 0.0648 & 0.0655 & 0.0657 & 0.0659 & 0.0113\\ 
0.015 & 0.085 & 0.0588 & 0.0671 & 0.0678 & 0.0682 & 0.0681 & 0.0132\\ 
0.010 & 0.090 & 0.0603 & 0.0681 & 0.0689 & 0.0692 & 0.0690 & 0.0146\\ 
0.005 & 0.095 & 0.0643 & 0.0698 & 0.0706 & 0.0710 & 0.0707 & 0.0143\\ 
\hline
\end{tabular}  
\end{table}

\begin{table}[th] 
\begin{tabular}{cccccccc}
\hline
$T_C^c$ & $T_C^h$ & $T_C^1$ & $T_C^2$ & $T_C^3$ & $T_C^4$ & $T_C^5$ & $Q$ \\   
\hline
0.0471 & 0.0537 & 0.0506 & 0.0506 & 0.0506 & 0.0506 & 0.0506 & 0.0020\\ 
0.0445 & 0.0578 & 0.0519 & 0.0522 & 0.0523 & 0.0522 & 0.0524 & 0.0039\\ 
0.0423 & 0.0623 & 0.0540 & 0.0548 & 0.0550 & 0.0550 & 0.0552 & 0.0057\\ 
0.0400 & 0.0672 & 0.0563 & 0.0578 & 0.0581 & 0.0581 & 0.0583 & 0.0076\\ 
0.0378 & 0.0723 & 0.0587 & 0.0608 & 0.0611 & 0.0610 & 0.0614 & 0.0094\\ 
0.0353 & 0.0775 & 0.0606 & 0.0631 & 0.0635 & 0.0635 & 0.0638 & 0.0113\\ 
0.0327 & 0.0831 & 0.0624 & 0.0655 & 0.0659 & 0.0659 & 0.0660 & 0.0132\\ 
0.0298 & 0.0886 & 0.0638 & 0.0669 & 0.0671 & 0.0670 & 0.0671 & 0.0146\\ 
0.0285 & 0.0937 & 0.0670 & 0.0694 & 0.0695 & 0.0695 & 0.0693 & 0.0143\\ 
\hline
\end{tabular} 

\end{table}

\newpage
\newpage

Table II.  Kinetic temperatures (above) and configurational temperatures
(below) are shown as functions of the long-time-averaged (one billion
timesteps) heat flux $Q$
induced by the temperature difference $T_K^h - T_K^c$ between two
thermostated Nos\'e-Hoover particles.  The first
seven columns correspond to the temperatures of the cold and hot
particles, followed by the temperature of the Newtonian particles
(the Newtonian particles are the five shown in a vertical column
in Fig. 3, and labeled from bottom to top.)
\begin{table}[th] 
\begin{tabular}{cccccccc}
\hline
$T_K^c$ & $T_K^h$ & $T_K^1$ & $T_K^2$ & $T_K^3$ & $T_K^4$ & $T_K^5$ & $Q$    \\   
\hline
0.001 & 0.003 & 0.00134 & 0.00146 & 0.00146 & 0.00146 & 0.00147 & 0.00002 \\ 
0.002 & 0.006 & 0.0029 & 0.0031 & 0.0031 & 0.0031 & 0.0031 & 0.00008 \\
0.005 & 0.015 & 0.0089 & 0.0100 & 0.0100 & 0.0100 & 0.0101 & 0.00064 \\
0.010 & 0.030 & 0.0207 & 0.0237 & 0.0238 & 0.0238 & 0.0242 & 0.00269 \\
0.020 & 0.060 & 0.0447 & 0.0504 & 0.0508 & 0.0508 & 0.0509 & 0.00736 \\
0.050 & 0.150 & 0.1066 & 0.1132 & 0.1142 & 0.1148 & 0.1152 & 0.01858 \\
\hline
\end{tabular}  
\end{table}

\begin{table}[th] 
\begin{tabular}{cccccccc}
\hline
$T_C^c$ & $T_C^h$ & $T_C^1$ & $T_C^2$ & $T_C^3$ & $T_C^4$ & $T_C^5$ &     $Q$ \\   
\hline
0.00125 & 0.00217 & 0.00135 & 0.00142 & 0.00140 & 0.00142 & 0.00145 & 0.00002 \\ 
0.0025  & 0.0043  & 0.0029 & 0.0030 & 0.0030 & 0.0030 & 0.0031 & 0.00008 \\
0.0075  & 0.0120  & 0.0095 & 0.0098 & 0.0097 & 0.0097 & 0.0099 & 0.00064 \\
0.0159  & 0.0265  & 0.0218 & 0.0229 & 0.0229 & 0.0229 & 0.0234 & 0.00269 \\
0.0311  & 0.0570  & 0.0470 & 0.0490 & 0.0492 & 0.0491 & 0.0494 & 0.00736 \\
0.0673  & 0.1497  & 0.1104 & 0.1125 & 0.1133 & 0.1136 & 0.1138 & 0.01858 \\
\hline 
\hline
\end{tabular}  
\end{table}

\end{document}